# Analysis of a superbolide from a Damocloid observed over Spain on July 13, 2012

José M. Madiedo[1,2], Josep M. Trigo-Rodríguez[3], Jaime Zamorano[4], José L. Ortiz[5], Alejandro Sánchez de Miguel[4], Francisco Ocaña[4], Jaime Izquierdo[4], Alberto J. Castro-Tirado[5], Nicolás Morales[5], David Galadí[6], Enrique de Guindos[6], Juan Lacruz[7], Faustino Organero[8], Leonor Ana-Hernández[8], Fernando Fonseca[8], Mar Tapia[9], Felipe Gallego[2] and Jesús Cabrera-Caño[2]

[1]Depto. de Física Atómica, Molecular y Nuclear, Facultad de Física, Universidad de Sevilla, 41012 Sevilla, Spain.
[2]Facultad de Ciencias Experimentales, Universidad de Huelva, Avda. de las Fuerzas Armadas S/N. 21071 Huelva, Spain.
[3]Institute of Space Sciences (CSIC-IEEC). Campus UAB, Facultat de Ciències, Torre C5-p2. 08193 Bellaterra, Spain.
[4]Depto. de Astrofísica y CC. de la Atmósfera, Facultad de Ciencias Físicas, Universidad Complutense de Madrid, 28040 Madrid, Spain.
[5]Instituto de Astrofísica de Andalucía, CSIC, Apt. 3004, 18080 Granada, Spain.
[6]Centro Astronómico Hispano-Alemán, Calar Alto (CSIC-MPG), E-04004 Almería, Spain.
[7]La Cañada Observatory (MPC J87), Ávila, Spain.
[8]Observatorio Astronómico de La Hita, La Puebla de Almoradiel, Toledo, Spain.
[9]Laboratori d'Estudis Geofísics Eduard Fontseré (LEGEF), IEC, Barcelona, Spain.

**ABSTRACT**

A superbolide with an estimated absolute magnitude of -20±1 was seen on July 13, 2012 over the center and south of Spain. This extraordinary event, which was witnessed by numerous casual observers, was recorded in the framework of the continuous fireball monitoring and meteor spectroscopy campaign performed by the SPanish Meteor Network (SPMN). Thus, because of optimal weather conditions, the bolide was imaged from ten meteor observing stations. Here we present the analysis of this magnificent event, which is the brightest fireball ever recorded by our team. The atmospheric trajectory of the bolide and the orbit in the Solar System of the parent meteoroid were obtained. The emission spectrum produced during the ablation of this particle is also discussed. We found that the meteoroid, which was following a Halley Type Comet orbit, was depleted in Na and had a tensile strength one order of magnitude higher than that corresponding to typical cometary materials. By means of orbital analysis tools we have investigated the likely parent body of this particle and the results suggest that the progenitor is a damocloid. The impact area of the hypothetical remnants of the meteoroid is also given and a search for meteorites was performed, but none were found.

**KEYWORDS:** meteorites, meteors, meteoroids, comets.

**1 INTRODUCTION**

At 0h04m52.6±0.1s UTC on July 13, 2012 an exceptionally bright fireball was observed across the center and south of the Iberian Peninsula. It was named "Belmonte de Tajo", since its maximum brightness was reached when the bolide was located next to the zenith of this town, at about 45 km southeast of Madrid. The event, which lasted around 3 seconds, was witnessed by numerous casual observers, mostly from their homes. Some of these reported a strong thunder-like sound heard a few seconds after the luminous phase of the event took place. Most witnesses also claimed that night suddenly turned into day during a fraction of a second. This, in fact, could be confirmed by our fireball monitoring systems. Thus, because of optimal weather conditions, the event was imaged from eight meteor observing stations operated by the SPanish Meteor Network and by one camera operating at the Calar Alto Astronomical Observatory (Table 1). Additional images were provided by the cameras operated by DGT (the Spanish Traffic Service). In this way, we could confirm that the event fell within the superbolide category (Ceplecha et al. 1999), since the photometric analysis of our video recordings showed that the bolide reached an absolute magnitude of -20±1 at its brightest phase. This fireball is, in addition, the brightest event ever recorded by our network as the also extraordinary Bejar bolide appeared on July 11, 2008 reached an absolute magnitude of -18 (Trigo-Rodríguez et al. 2009). Like in this previous case, the opportunity of reconstructing the trajectory, radiant and orbital elements of the July 13, 2012 superbolide proves to be crucial to better understand the origin and dynamics of meter-sized meteoroids orbiting the Sun. We also present the bulk chemistry of the meteoroid producing this event from the analysis of its emission spectrum.

## 2 INSTRUMENTATION AND METHODS

The fireball was imaged from eight SPMN meteor observing stations: Sevilla, La Hita, Huelva, El Arenosillo, Madrid, Villaverde del Ducado, Sierra Nevada and Ávila. It was also recorded by a low-resolution video camera operating at Calar Alto. The geographical coordinates of these observing stations are given in Table 1. SPMN systems are based on high-sensitivity monochrome CCD video devices we employ to monitor the night sky (models 902H and 902H Ultimate, from Watec Co.). These can image meteor events with an apparent magnitude of around +3/+4, or brighter. Station #1 also employs colour CCD video cameras. Although these were set up to monitor the sky during the day, we usually keep them working also during the night to image events with brightness higher than that corresponding to an apparent magnitude of about -4. All of these CCD devices (monochrome and colour) operate according to the PAL video standard. With this configuration, these cameras generate interlaced video sequences with a frame rate of 25 fps and a resolution of 720x576 pixels (Trigo-Rodriguez et al. 2007; Madiedo & Trigo-Rodríguez 2008; Madiedo et al. 2010).

Some of our observing stations perform a continuous spectroscopic monitoring since 2006 to infer information about the chemical composition of meteoroids. For this purpose we employ holographic diffraction gratings (500 or 1000 lines/mm, depending on the device) attached to the objective lens of some of the above-mentioned cameras to image the emission spectra resulting from the ablation of meteoroids in the atmosphere (Madiedo et al. 2013a, 2013b, 2013c). Our video spectrographs can image spectra for meteor events with brightness higher than the corresponding to an apparent magnitude of around -4.

With respect to data reduction, we first deinterlaced the video sequences by separating the even and odd fields contained in each video frame. Then, an astrometric measurement was done by hand in order to obtain the plate (x, y) coordinates of the fireball along its apparent path in the sky from each observing station. These astrometric measurements were then introduced in the AMALTHEA software (Madiedo et al. 2011a), which transforms plate coordinates into equatorial coordinates by using the position of reference stars appearing in the images. This follows the method of the intersection of planes to obtain the position of the apparent radiant and the trajectory in the atmosphere of multi-station meteors (Ceplecha 1987). Once these data are known, the software computes the orbital parameters of the corresponding meteoroid by following the procedure described in Ceplecha (1987).

On the other hand, the magnitude of the fireball was estimated from the photometric analysis of the images recorded by our CCD video cameras. In general, we determine the magnitude of meteor events by direct comparison of the brightness level of the pixels along the meteor trail and those corresponding to the position of nearby stars. However, since the event discussed here saturated the video frames containing its brightest phase, a different approach was employed. Thus, we employed the images taken by other cameras that did not contain the fireball in their field of view but, however, did record the increase in brightness of the environment as a consequence of the luminosity of the bolide. This brightness level was then compared to a previous calibration performed by us where the luminosity of this environment as a consequence of the different phases of the Moon during the night but also of the Sun at dawn and sunset was measured. Then, by using this approach, the peak magnitude of this event was obtained with an uncertainty of about ±1.

**3 OBSERVATIONS: atmospheric path, radiant and orbit**

The "Belmonte de Tajo" superbolide was included in our fireball database with the SPMN code 130712. This code was given after the observing date. Because of its extraordinary brightness, it was witnessed by many casual observers along Spain. In fact, our video fireball monitoring cameras operating at the Sierra Nevada Astronomical Observatory (Granada, Spain) recorded the shadows produced by the fireball at over 300 km from the epicentre of this event. Figures 1a to 1f show several images of this bolide as recorded from observing stations #2 and #7 during the initial phase, half position in the atmospheric path and maximum brightness of the fireball. The image taken by the low-resolution camera operating at Calar Alto is shown in Figure 1h.

Some of the witnesses reported their observations to the SPMN website, via an online form that is available for these purposes. Most of these casual observers commented that the fireball had a greenish colour. This was confirmed by one of our fireball monitoring cameras operating from Sevilla, since this device employs a colour CCD sensor. These reports, however, did not contain accurate information about the atmospheric trajectory of the bolide and so they were not used for the orbit determination. So, the analysis of its trajectory and orbit was performed on the basis of the images recorded from SPMN video stations.

The calculated parameters of the atmospheric trajectory of the superbolide are summarized in Table 2. The preatmospheric velocity $V_\infty$ was found from the velocities measured at the earliest part of the meteor trajectory. No deceleration was observed between the initial height and ~70 km. Taking this into account, the parent meteoroid struck the atmosphere with an initial velocity $V_\infty=22.6\pm0.3$ km s$^{-1}$. The fireball started its luminous path at 100.4±0.5 km above the ground level and reached its maximum brightness (corresponding to an absolute magnitude of -20±1) at a height of 36.7±0.5 km. The luminous phase ended at 35.4±0.5 km. The inclination of this trajectory was of about 75º with respect to the ground and the position of the apparent radiant was $\alpha=261.5\pm0.1°$, $\delta=37.66\pm0.06°$. With this information the heliocentric orbit in the Solar System of the parent meteoroid was calculated. The corresponding orbital parameters (J2000) are listed in Table 3 and the projection of this orbit on the ecliptic plane is shown in Figure 1i. The calculated aphelion is Q=14±2 AU and the revolution period yields P=20.1±4.0 yr, with a Tisserand parameter with respect to Jupiter $T_J=1.8\pm0.1$. Thus, although these data satisfy the conditions imposed to Halley Type Comet (HTC) orbits (P>20yr and $T_J$<2), the orbit of the meteoroid can be regarded as a limiting case between those of Jupiter Family (JFC) and HTC.

## 4 DISCUSSION
### 4.1 Light curve and preatmospheric mass

A photometric analysis was performed on the video frames containing the fireball. From this we have obtained the evolution with time of the luminosity of this deep-penetrating bolide. The resulting light curve is plotted in Figure 4. This reveals that the fireball experienced several bright flares during the second half of its atmospheric trajectory, between t=1.6 and t=2.6 seconds. This behaviour is exhibited by compact meteoroids (Murray et al. 1999; Campbell et al. 2000), while for dustball meteoroids these flares take place earlier (Hawkes & Jones 1975). The last of these events, which is also the brightest one, corresponds to the violent disruption of the meteoroid at the end of the luminous trajectory.

The light curve has been employed to infer the initial (preatmospheric) mass of the meteoroid. For this purpose, the classical meteor luminous equation has been used:

$$m = 2 \int_{t_e}^{t_b} I/(\tau v^2)\, dt \quad (1)$$

Here I is the measured luminosity of the fireball, which is given as a function of the absolute magnitude M by means of the relationship

$$I = 10^{-0.4 \cdot M} \quad (2)$$

For the calculation of the luminous efficiency ($\tau$), which depends on velocity (v) we have employed the equations given by Ceplecha & McCrosky (1976). In this way, the initial mass of the meteoroid yields 1700±500 kg. According to this, the diameter of this particle would range from 1.1±0.1 m to 0.9±0.1 m for a bulk density of 2.4 g cm$^{-3}$ and 3.7 g cm$^{-3}$, respectively. Since even meter-sized meteoroids produced by comets are completely disintegrated at high altitudes in the atmosphere (Madiedo et al. 2013d), the deep penetration ability of the "Belmonte de Tajo" meteoroid indicates that this particle was tougher than ordinary cometary meteoroids.

### 4.2 Tensile strength

The compactness of the meteoroid inferred from the photometric behaviour of the fireball can be confirmed by estimating the tensile strength of the particle. Thus, the bolide exhibited several flares along its atmospheric trajectory as a consequence of the successive break-ups suffered by the meteoroid. These can be noticed in the light curve plotted in Figure 2. However, the violent disruption of the meteoroid took place by the end of the luminous phase where the major flare was produced and the event reached its maximum brightness. Previous flares were not catastrophic, since the meteoroid continued penetrating in the atmosphere. The equation given by Bronshten (1981) can be employed to estimate the tensile (aerodynamic) strength S at which these flares took place:

$$S = \rho_{atm} \cdot v^2 \qquad (3)$$

This aerodynamic strength S can be used as an estimation of the tensile strength of the meteoroid (Trigo-Rodriguez & Llorca 2006, 2007). In this relationship v is the velocity of the meteoroid at the break-up point and $\rho_{atm}$ is the atmospheric density at the height where the flare takes place. We have calculated this density by means of the US standard atmosphere model (U.S. Standard Atmosphere 1976). Table 4 shows the values of the aerodynamic strength for the different flares highlighted in Figure 2. The corresponding heights and velocities are also indicated. From this analysis we found that the value obtained for the final flare, $(3.1\pm0.6)\cdot10^7$ dyn cm$^{-2}$, is 2 orders of magnitude higher than the typical tensile strengths of $\sim10^5$ dyn cm$^{-2}$ exhibited by tough meteoroids with a cometary origin (Trigo-Rodriguez and Llorca 2006, 2007), but similar to those found for stony meteoroids (Consolmagno and Britt 1998; Consolmagno et al. 2006, 2008; Macke et al. 2011).

**4.3 Analysis of the emission spectrum**

Three video spectrographs operating from stations #2, #5 and #6 in Table 1 recorded the emission spectrum produced during the ablation of the meteoroid in the atmosphere. We have employed our CHIMET software to process these spectra (Madiedo et al. 2011b). Thus, the video files containing the emission spectrum were first deinterlaced by separating even and odd video fields, so that a new video sequence was produced from them. Then, the resulting video file was dark-frame substracted and flat-fielded. A calibration in wavelengths was then performed with by identifying typical lines appearing in meteor spectra and, finally, the signal was corrected by taking into account the spectral efficiency of the spectrograph. The resulting spectrum is shown in Figure 3, where multiplet numbers are given according to Moore (1945). As can be seen, the signal is dominated by a strong emission from Mg I-2 (517.2 nm) and several Fe I multiplets: Fe I-41 (441.5 nm) and Fe I-15 (526.9 and 542.9 nm). In the ultraviolet, the intensity of H and K lines of ionized calcium is also strong. The contribution from Na I-1 is, however, very small, revealing a depletion of sodium in the meteoroid. The emission from atmospheric $N_2$ bands can also be noticed in the red region.

Once the main emission lines were identified, the relative abundances were calculated. Thus, a software application was used to reconstruct a synthetic spectrum by adjusting the temperature (T) in the meteor plasma, the column density of atoms (N), the damping constant (D), and the surface area (P) from the observed brightness of lines as explained in Trigo-Rodríguez et al. (2003). First, since Fe I emission lines are well distributed all over the spectrum, they were used to set the parameters capable to fit the observed spectrum with the synthetic one (Trigo-Rodríguez et al. 2003, 2004). The modelled spectrum of the meteor column is then compared line by line with the observed one to fit all its peculiarities. Thus, once the parameters T, D, P and N are fixed, the relative abundances of the main rocky elements relative to Fe are found by performing an iterative process until an optimal fit is achieved. To obtain the elemental abundances relative to Si, we have considered Fe/Si=1.16 (Anders and Grevesse 1989). The results are summarized in Table 5, where these relative abundances are compared with those found for other undifferentiated bodies in the Solar System, such as comet 1P/Halley and the CI and CM groups of carbonaceous chondrites (Jessberger et al. 1988; Rietmeijer & Nuth 2000; Rietmeijer 2002; Trigo-Rodríguez et al. 2003). Overall, the computed abundances suggest that the main rock-forming elements were available in proportions close to those found for comet 1P/Halley, but some small differences suggest that perhaps we were in front of not yet sampled solar system materials. In particular, the SPMN130712 meteoroid exhibited a very significant depletion in Na.

**4.4 Orbital analysis: parent body**

The results obtained from the analysis of the photometric behaviour and the emission spectrum of the "Belmonte de Tajo" superbolide show the existence of tough meteoroids depleted in Na moving in cometary orbits. We have employed the orbital elements inferred from the analysis of this superbolide to find the likely parent body of this meteoroid among the objects currently listed in the NeoDys (http://newton.dm.unipi.it/neodys/) and the Minor Planet Center (http://www.minorplanetcenter.org/iau/mpc.html) databases. One of the advantages of finding the parent body is the possibility to establish a correlation with the physical and chemical properties of the meteoroid and those of its progenitor. For this purpose, we have employed the ORAS software (ORbital Association Software), which can perform a link between meteoroids and their potential progenitor bodies on the basis of dissimilarity criteria (a review about this topic is given in Williams (2011)). In this way, the value of the so-called dissimilarity function, which measures the "distance" between the orbit of the meteoroid and that of the potential parent body, is obtained. Thus, a link can be established if this distance is below a predefined cut-off value. This software, which is described in detail in Madiedo et al. (2013a), can employ the different dissimilarity criteria proposed by Southworth and Hawkins (1963), Drummond (1981), Jopek (1993), Valsecchi et al. (1999) and Jenniskens (2008). However, no parent body was found, which means that either the progenitor is not yet catalogued or the link with the parent body is broken because of the differences in orbital elements arisen during the evolution of the orbit of the progenitor and the meteoroid (Vaubaillon et al. 2006).

Since the specific parent body could not be determined, we have employed a different approach in order to infer more information about the nature and past history of the progenitor and, in this way, try to explain some of the physical properties calculated for the SPMN130712 meteoroid. Thus, we have analyzed the evolution of the orbit of this particle by integrating backwards in time its orbital elements by means of the Mercury 6 symplectic integrator (Chambers 1999). In fact, a lower perihelion at an earlier epoch could explain the inferred Na depletion and high tensile strength on the basis of the sublimation of volatiles. This situation is found, for instance, for Geminid meteoroids, where their low volatile content is thought to be due to the low perihelion distance ($q=0.14$ AU) of this stream (Čapek & Borovička 2009). Besides, meteoroids with $q<0.1$ (as, for instance, the δ-Aquariids) are found to have Na-free properties (Kasuga et al. 2006).

For this calculation, the gravitational fields of Venus, the Earth-Moon system, Mars, Jupiter, Saturn, Uranus, Neptune and Pluto were considered. In order to take into account the uncertainty in the orbital elements of the meteoroid, a set of 100 clones was created around the orbit of this particle. Their orbital elements were spread within the error bars determined for the orbit of the meteoroid and at the 3-sigma confidence interval. The orbits were integrated back for $5 \cdot 10^5$ years. This analysis reveals that only three clones exhibited a decrease of the perihelion distance below 0.14 AU after a time period ranging between 45,000 and 80,000 years, reaching values below 0.03 AU. Because of thermal desorption of alkali silicates, at such small distances from the Sun mm-sized meteoroids are depleted in Na, but also in Fe and Mg as a consequence of the sublimation of additional minerals containing these elements (Kasuga et al. 2006).

However, this should be taken with caution here, as we could expect that for meter-sized meteoroids such as SPMN130712 thermal desorption is not so effective. Nevertheless, if we admit that thermal desorption at low perihelion distances was able of sublimating Na containing minerals in the SPMN130712 meteoroid, we should also admit that such mechanism must decrease, at least to a significant extent, the Fe and Mg contents. Since these additional depletions are not observed in the emission spectrum, we conclude that the orbits of clones exhibiting a small perihelion distance, which are also minority, do not account for the physical properties inferred for the SPMN130712 meteoroid. On the contrary, most clones (92 out of 100) exhibited an increase of the semimajor axis and orbital eccentricity. As a consequence of this, the aphelion of these clones was placed between 90 and 130 AU after a time period ranging between 50,000 and 440,000 years in the past. The perihelion distance also increased or remained of the order of 1 AU for these of clones. On dynamical grounds, these orbits correspond to either Halley Type Comets or Long Period Comets, but as previously mentioned the strength of cometary materials is much lower than the tensile strength inferred for the SPMN130712 meteoroid (Trigo-Rodriguez & Llorca 2006, 2007; Trigo-Rodríguez and Blum 2009). There are, however, other minor bodies in the Solar System which are known to follow HTC orbits: the damocloids. The Tisserand parameter with respect to Jupiter of the damocloids, whose archetypal object is 5335 Damocles, is not larger than 2 (Jewitt 2005). According to the JPL Small Body Database (http://neo.jpl.nasa.gov/), only 88 minor bodies are currently identified as damocloid candidates, although most of them have poorly defined orbits (with an arc-span below 30 days). The fact that the SPMN130712 meteoroid originated from a damocloid would explain the low Na content and high tensile strength determined from the analysis of the "Belmonte de Tajo" superbolide. Thus, the damocloids are considered to be nuclei of dead Halley-family and long-period comets and, so, they are depleted in volatiles (Jewitt 2005; Jewitt et al. 2008; Fernández et al. 2005; Toth 2006). According to an alternative scenario proposed to explain the origin of the damocloids, the source of some of these bodies is probably associated with the sweeping resonances found in the Nice model (Gomes et al. 2005; Morbidelli et al. 2005, 2010). The large movement of the giant planets about 4.1 to 3.9 Gyrs ago caused powerful orbital resonances to sweep across the asteroid belt, causing a significant mass depletion of bodies in the belt (Bottke et al. 2012; Walsh et al. 2012). In this way, the dynamic evolution and disruption of asteroids in these unusual high-inclination orbits could be the source of the damocloids, but the present population of these asteroids moving in HTC orbits could be a minor remnant of its importance in the past due to the fact that they were mostly destined to bombard the terrestrial planets long time ago (Morbidelli et al. 2010).

**4.5 Dark flight and field search**

Because of the size of the event and the penetration of the fireball till a final altitude of about 35 km, the survival of fragments that could be recovered as meteorites on the ground was not dismissed. So, the dark flight of these fragments was modelled by means of the AMALTHEA software, which performs this analysis by following the procedure described in (Ceplecha 1987). Atmospheric data provided by AEMET (the Spanish meteorological agency) were used to include wind effects in the calculations. Because of the violent disruption of the meteoroid by the end of the luminous phase, we expected that most of the surviving fragments would be small, with a mass below several tens of grams. Since the inclination of the trajectory with respect to the ground was of about 75º, these, however, would be dispersed over a relatively small area on the ground. The shape of these fragments was supposed to be spherical and a value of the drag coefficient $\Gamma=0.51$ was considered. According to this, the landing point would be located around the geographical coordinates 40.1º N, 3.2º W. Yet, despite several expeditions were organized to the strewnfield, no meteorites were found.

**5 CONCLUSIONS**

We have analyzed a mag. -20±1 superbolide observed over Spain on July 13, 2012. Our main conclusions are listed below:

1) The meteoroid, with an estimated initial mass of 1,700±500 kg and a diameter of about 1 meter, penetrated the atmosphere till a final height of 35.4±0.5 above the ground level. By that point, it experienced a very intense flare as a consequence of the sudden disruption of the body.
2) The tensile strength of the meteoroid was estimated by calculating the aerodynamic pressure at the disruption point. The result supports the existence of high-strength meteoroids moving in cometary orbits.
3) The analysis of the emission spectrum produced by the fireball provided the elemental abundances of the main rocky elements in the meteoroid. This spectrum indicates that the meteoroid was quite poor in Mg and exhibited a very significant depletion in Na. Overall, the computed abundances suggest that the main rock-forming elements were available in proportions close to those found for comet 1P/Halley, but some small differences might suggest that we were in front of still non sampled solar system materials.

4) The specific parent body of the meteoroid could not be established by performing a search through the minor bodies currently listed in the NeoDys and Minor Planet Center databases. However, the integration backwards in time of the orbital elements of the meteoroid suggests that the parent body is a damocloid. So, the analysis of the "Belmonte de Tajo" fireball has provided an insight into the physical properties of dead nuclei of HTCs.

5) Because of the size of the event and the terminal height of the luminous trajectory, the survival of fragments as meteorites was considered and their landing point was calculated. Several expeditions to the expected impact area were organized, although no meteorites were found.

**ACKNOWLEDGEMENTS**

We thank *AstroHita Foundation* for its support in the establishment and operation of the automated meteor observing station located at La Hita Astronomical Observatory (La Puebla de Almoradiel, Toledo, Spain). We also acknowledge support from the Spanish Ministry of Science and Innovation (projects AYA2009-13227, AYA2009-06330-E and AYA2011-26522) and Junta de Andalucía (project P09-FQM-4555). We also thank *Dirección General de Tráfico* for providing additional images of the event discussed in the text.

**REFERENCES**

Anders E., Grevesse N., 1989, Geochimica et Cosmochimica Acta, 53, 197.

Bottke W.F. et al., 2012, Nature, 458, 78.

Bronshten V.A., 1981, Geophysics and Astrophysics Monographs. Reidel, Dordrecht.

Campbell M.D. et al., 2000, Meteorit. & Planet. Sci., 35, 1259.

Čapek D., Borovička J., 2009, Icarus, 202, 361.

Ceplecha Z., 1987, Bull. Astron. Inst. Cz., 38, 222.

Ceplecha Z. et al., 1999, Astron. Inst. Slovak Acad. Sci., Bratislava, 37-54.

Ceplecha Z. & McCrosky R.E., 1976, Journal of Geophysical Research, 81, 6257.

Consolmagno G.J. and Britt D.T., 1998, Meteoritics & Planetary Science, 33, 1231.

Consolmagno G.J., Macke R.J., Rochette P., Britt D.T., and Gattacceca J., 2006, Meteoritics & Planetary Science, 41, 331.

Consolmagno G.J., Britt D.T., and Macke R.J., 2008, Chemie der Erde-Geochemistry, 68, 1.

Chambers J.E., 1999, MNRAS, 304, 793.

Drummond J.D., 1981, Icarus, 45, 545.

Fernández Y.R, Jewitt D., Sheppard S.S., 2005, AJ, 130, 308.

Gomes R., Levison H.F., Tsiganis K., Morbidelli A., 2005, Nature, 435, 466.

Hawkes R.L. & Jones J., 1975, MNRAS, 173, 339.

Jenniskens P., 2008, Icarus, 194, 13.

Jewitt D., 2005, AJ, 129, 530.

Jewitt D., Morbidelli A., Rauer H., 2008, Trans-Neptunian Objects and Comets, Springer.

Jessberger E.K., Christoforidis A., Kissel J., 1988, Nature, 332, 691.

Jopek T.J., 1993, Icarus, 106, 603.

Kasuga T., Yamamoto T., Kimura H., Watanabe J., 2006, A&A, 453, L17.

Macke R.J., Consolmagno G.J., and Britt D.T. et al., 2011, Meteoritics & Planetary Science, 46, 1842.

Madiedo J.M., Trigo-Rodríguez J.M., 2008, Earth Moon Planets, 102, 133.

Madiedo J.M., Trigo-Rodríguez J.M., Ortiz J.L., Morales N., 2010, Advances in Astronomy, 2010, 1.

Madiedo J.M., Trigo-Rodríguez J.M., Lyytinen E., 2011a, NASA/CP-2011-216469, 330.

Madiedo J.M., Trigo-Rodríguez, J.M., Williams I.P., Ortiz J.L., Cabrera J., 2013a, MNRAS, 431, 2464.

Madiedo J.M., Trigo-Rodríguez, J.M., Konovalova N., Williams I.P., Castro-Tirado A.J., Ortiz J.L., Cabrera J., 2013b, MNRAS, 433, 571.

Madiedo J.M., Trigo-Rodríguez J.M., Lyytinen E., Dergham J., Pujols P., Ortiz J.L., Cabrera J., 2013c, MNRAS, 431, 1678.

Madiedo J.M. et al., 2013d, Icarus, submitted.

Moore C.E., 1945, In: A Multiplet Table of Astrophysical Interest. Princeton University Observatory, Princeton, NJ. Contribution No. 20.

Morbidelli A. et al., 2005, Nature, 435, 462.

Morbidelli A. et al., 2010, Astron. J., 140, 1391.

Murray I.S., Hawkes R.L., Jenniskens P., 1999, Meteorit. & Planet. Sci., 34, 949.

Rietmeijer F. and Nuth J.A., 2000, Earth, Moon and Planets, 82-83, 325.

Rietmeijer F., 2002, Chemie der Erde, 62, 1.

Southworth R.B., Hawkins G.S., 1963, Smithson Contr. Astrophys., 7, 261.

Toth I., 2006, Proceedings of IAU Symposium No. 229, 67.

Trigo-Rodríguez J.M., Llorca J., 2007, Adv. Space Res., 29, 517.

Trigo-Rodríguez J.M., Blum J., 2009, Plan. Space Sci., 57, 243.

Trigo-Rodríguez J.M., Llorca J., Borovicka J., Fabregat J., 2003, Meteorit. Planet. Sci., 38, 1283.

Trigo-Rodríguez J.M., Llorca J., 2006, MNRAS, 372, 655.

Trigo-Rodríguez J.M., Llorca J., 2007, MNRAS, 375, 415.

Trigo-Rodríguez J.M. et al., 2007, MNRAS, 382, 1933.

Trigo-Rodríguez J.M., Madiedo J.M., Williams I.P., et al. 2009, MNRAS, 394, 569.

U.S. Standard Atmosphere, 1976, NOA-NASA-USAF, Washington.

Valsecchi G., Jopek T., Froeschlé C., 1999, MNRAS, 304, 743.

Vaubaillon J., Lamy P., Jorda L., 2006, MNRAS, 370, 1841.

Walsh K.J. et al., 2012, MAPS, 47, 1941.

Williams I.P., 2011, A&G, 52, 2.20.

**TABLES**

Table 1. Geographical coordinates of the meteor observing stations involved in this work.

| Station # | Station name | Longitude (W) | Latitude (N) | Alt. (m) |
|---|---|---|---|---|
| 1 | Sevilla | 05º 58' 50" | 37º 20' 46" | 28 |
| 2 | La Hita | 03º 11' 00" | 39º 34' 06" | 674 |
| 3 | Huelva | 06º 56' 11" | 37º 15' 10" | 25 |
| 4 | Sierra Nevada | 03º 23' 05" | 37º 03' 51" | 2896 |
| 5 | El Arenosillo | 06º 43' 58" | 37º 06' 16" | 40 |
| 6 | Villaverde del Ducado | 02º 29' 29" | 41º 00' 04" | 1100 |
| 7 | Madrid-UCM | 03º 43' 34" | 40º 27' 03" | 640 |
| 8 | Ávila | 04º 29' 30" | 40º 36' 18" | 1400 |

Table 2. Trajectory and radiant data for the SPMN130712 fireball (J2000).

| $H_b$ (km) | $H_e$ (km) | $\alpha_g$ (º) | $\delta_g$ (º) | $V_\infty$ (km/s) | $V_g$ (km/s) | $V_h$ (km/s) |
|---|---|---|---|---|---|---|
| 100.4±0.5 | 35.4±0.5 | 258.5±0.1 | 36.98±0.07 | 22.6±0.3 | 19.8±0.3 | 40.3±0.3 |

Table 3. Orbital parameters (J2000) for the SPMN130712 fireball.

| a (AU) | e | i (º) | ω (º) | Ω (º) |
|---|---|---|---|---|
| 7.4±0.9 | 0.87±0.02 | 25.6±0.3 | 204.1±0.1 | 110.8413±10$^{-4}$ |

Table 4. Aerodynamic pressure for flares and break-up processes highlighted in Figure 3.

| Flare # | Height (km) | Velocity (km s$^{-1}$) | Aerodynamic pressure (dyn cm$^{-2}$) |
|---|---|---|---|
| 1 | 56±1 | 22.3±0.6 | (2.3±0.6)·10$^6$ |
| 2 | 55±1 | 22.3±0.6 | (2.6±0.6)·10$^6$ |
| 3 | 49±1 | 22.0±0.6 | (5.3±0.6)·10$^6$ |
| 4 | 45±1 | 21.7±0.6 | (8.8±0.6)·10$^6$ |
| 5 | 36±1 | 21.0±0.6 | (3.1±0.6)·10$^7$ |

Table 5. Derived elemental abundances relative to Si of the SPMN130712 superbolide compared with the inferred for other Solar System undifferentiated materials (Jessberger et al. 1988; Rietmeijer & Nuth 2000; Rietmeijer 2002).

| | Mg | Na | Ca | Mn ($\times 10^{-4}$) | Ni ($\times 10^{-3}$) | T (K) |
|---|---|---|---|---|---|---|
| SPMN130712 | 0.48 | 0.003 | 0.025 | 40 | 41 | 4700 |
| 1P/Halley | 0.54 | 0.054 | 0.034 | 30 | 22 | - |
| IDPs | 0.85 | 0.085 | 0.048 | 150 | 37 | - |
| CI chondrites | 1.06 | 0.060 | 0.071 | 90 | 51 | - |
| CM chondrites | 1.04 | 0.035 | 0.072 | 60 | 46 | - |

**FIGURES**

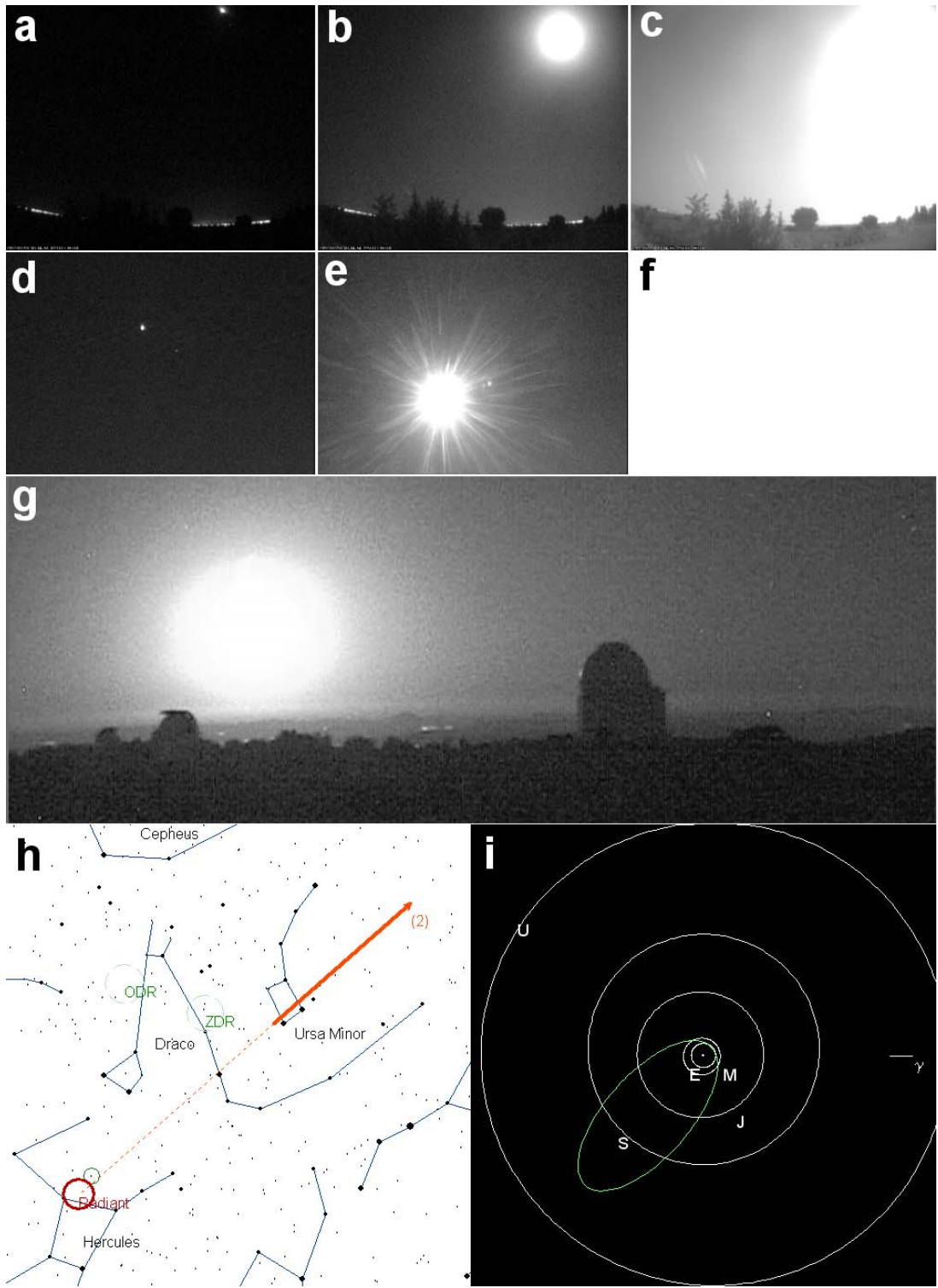

Figure 1. Video frames showing the initial phase, half position in the atmospheric path and maximum brightness of the fireball as imaged from La Hita Astronomical Observatory - a), b) and c) - and Madrid - d), e) and f) -. g) Image taken from Calar Alto. h) Apparent trajectory in the sky as seen from La Hita. i) Projection on the ecliptic plane of the heliocentric orbit of the meteoroid.

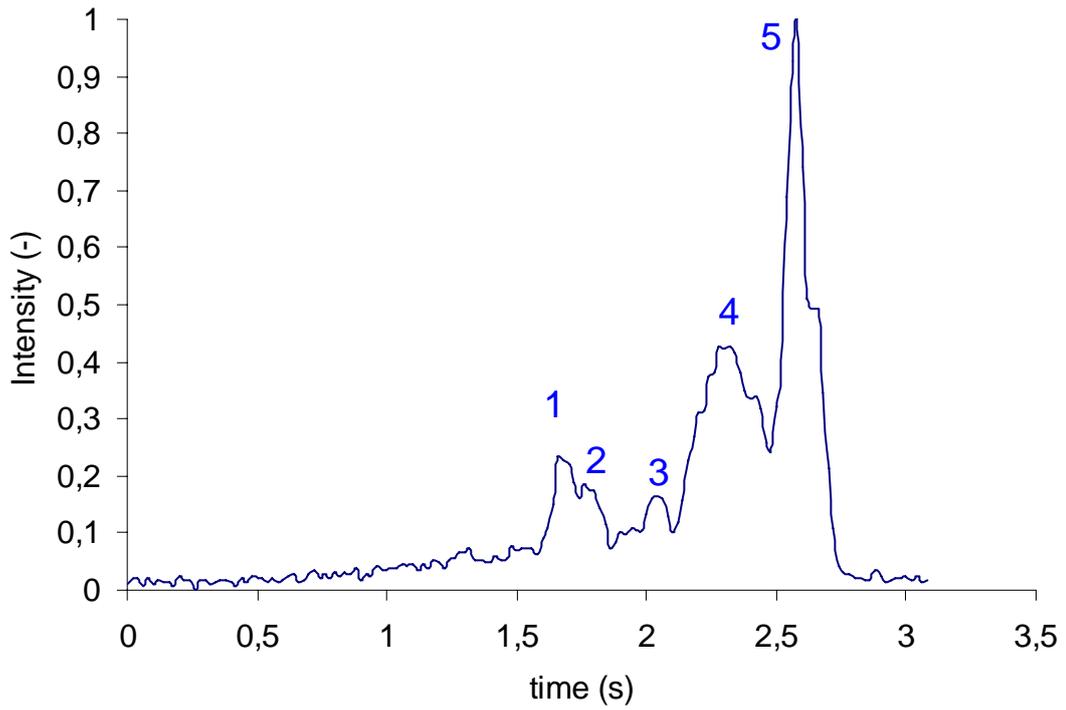

Figure 2. Light curve (relative pixel intensity vs. time) of the SPMN130712 fireball.

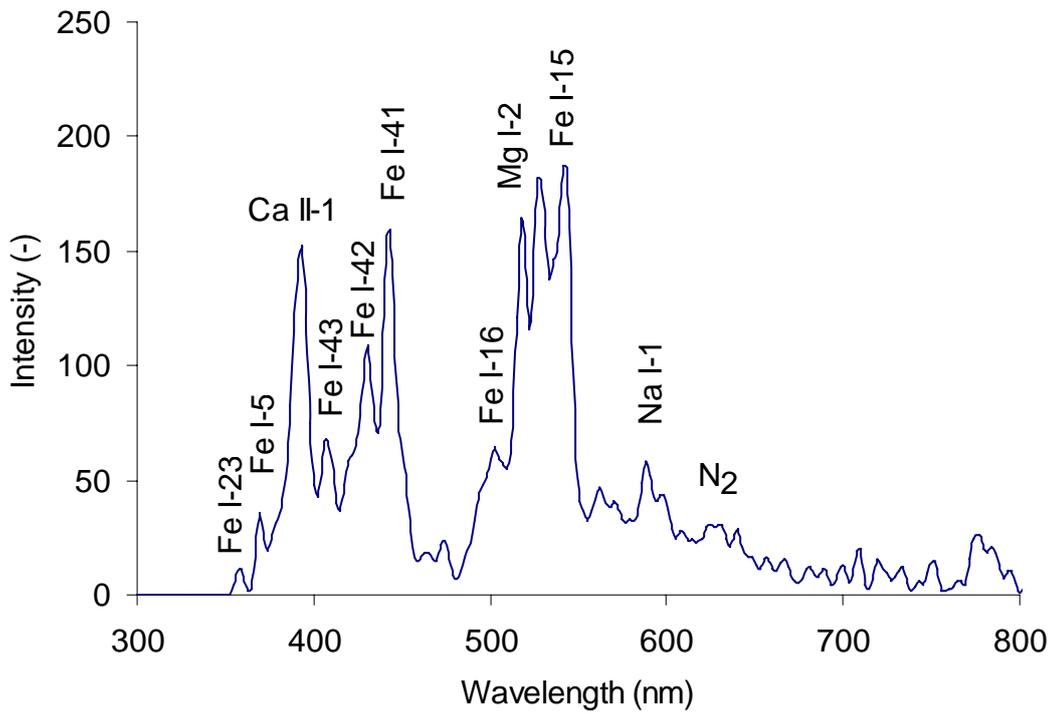

Figure 3. Calibrated emission spectrum of the SPMN130712 superbolide. Intensity is expressed in arbitrary units.

19